\documentclass[12pt]{article}
\usepackage{amssymb}
\usepackage{amstext}
\usepackage{amsmath}
\usepackage{latexsym}
\usepackage{color}

\newcommand{\R}{\mathbb R}

\newcommand{\mc}{\mathcal}

\newcommand{\be}{\begin{equation}}
\newcommand{\en}{\end{equation}}
\newcommand{\D}{{\mc D}}

\newcommand{\Lc}{{\cal L}}

\newtheorem{thm}{Theorem}
\newtheorem{cor}[thm]{Corollary}

\newtheorem{defi}{Definition}[section]
\newtheorem{lem}[defi]{Lemma}
\newtheorem{theorem}[defi]{Theorem}

\newcommand{\bedefin}{\begin{defi}}
\newcommand{\findefi}{\end{defi} \medskip}

\newcommand{\betheo}{\begin{theorem}$\!\!${\bf \,\,\,}}
\newcommand{\entheo}{\end{theorem}}
\newcommand{\enth}{\end{theorem}}

\newcommand{\becor}{\begin{cor}$\!\!${\bf .}}
\newcommand{\encor}{\end{cor}}

\newcommand{\belem}{\begin{lem}$\!\!${\bf .}}
\newcommand{\enlem}{\end{lem}}

\newcommand{\bea}{\begin{eqnarray}}
\newcommand{\ena}{\end{eqnarray}}

\newcommand{\beano}{\begin{eqnarray*}}
\newcommand{\enano}{\end{eqnarray*}}

\newcommand{\bee}{\begin{enumerate}}
\newcommand{\ene}{\end{enumerate}}

\newcommand{\bei}{\begin{itemize}}
\newcommand{\eni}{\end{itemize}}

\newcommand{\betab}{\begin{tabular}}
\newcommand{\entab}{\end{tabular}}

\newcommand{\bd}{\begin{displaymath}}

\newcommand{\h}{{\mathfrak H}}

\newcommand{\p}{\mbox{${\mathcal P}^{\uparrow}_{+}(1,1)$}}

\newcommand{\bpsi}{\mbox{\boldmath $\psi$}}

\newcommand{\bvarepsilon}{\mbox{\boldmath $\varepsilon$}}

\newcommand{\bxi}{\mbox{\boldmath $\xi$}}

\newcommand{\bPhi}{\mbox{\boldmath $\Phi$}}
\newcommand{\bPsi}{\mbox{\boldmath $\Psi$}}

\catcode `\@=11 \@addtoreset{equation}{section}

\catcode `\@=12

\begin{document}

\begin{center}
{\Large \bf SUSY Associated Vector Coherent States and Generalized
Landau Levels Arising From 2-dimensional
SUSY}   \vspace{2cm}\\

{\large S. Twareque Ali} \footnote[1]{Department of Mathematics and
Statistics, Concordia University,
Montr\'eal, Qu\'ebec, CANADA H3G 1M8\\
e-mail: stali@mathstat.concordia.ca}
\vspace{3mm}\\

{\large F. Bagarello} \footnote[2]{ Dipartimento di Metodi e Modelli
Matematici,
Facolt\`a di Ingegneria, Universit\`a di Palermo, I-90128  Palermo, ITALY\\
e-mail: bagarell@unipa.it\,\,\,\, Home page:
www.unipa.it/$^\sim$bagarell}
\end{center}

\vspace*{2cm}

\parskip=6pt

\begin{abstract}
\noindent We describe a method for constructing vector coherent
states for  quantum supersymmetric partner Hamiltonians. The method
is then applied to such partner Hamiltonians arising from a
generalization of the fractional quantum Hall effect. Explicit
examples are worked out.

\end{abstract}

\vfill

\newpage

% Section 1
\section{Introduction}

   Two quantum mechanical problems are addressed in this paper. The first is the
construction of vector coherent states, associated to supersymmetric
pairs of Hamiltonians and the second is a generalization of the
concept of Landau levels in the fractional quantum Hall effect, via
supersymmetric pairs of Hamiltonians. Coherent states, in the
context of supersymmetric quantum mechanics, have been studied
before (see for, example, \cite{{fatkosnietru},{ferhunie},{juroy}}).
These attempts were mainly centered around  building such states
from the eigenvectors of the Hamiltonian for the fermionic sector,
exploiting the {\em trilinear lowering operator} that can be
constructed using these vectors.  In this paper we adopt a different
strategy, in that we build {\em vector coherent states} using the
eigenvectors and eigenvalues of the pair of supersymmetric
Hamiltonians. This gives us coherent states which represent both the
bosonic and fermionic sectors. Our construction also makes contact
with another suggestion, that has recently been made in the
literature, which explicitly uses anti-commuting Grassmann variables
\cite{{borklilesri},{loaup}}, to introduce a quantization using
super Toeplitz operators.

The  so-called Landau levels appear in the analysis of the quantum
motion of an electron in a uniform magnetic field. This, in turn, is
the building block of a fascinating problem in many-body theory, the
{\em quantum Hall effect} (QHE), (see \cite{b6} and references
therein). We will not discuss here the role of these Landau levels
in the context of the QHE, which have been analyzed in many papers
and textbooks. Rather, we shall show how the use of two-dimensional
sypersymmetry (2d-SUSY), as discussed in \cite{daspern}, can be
useful to construct different super-partner Hamiltonians which, in
many  ways, behave analogously to the Hamiltonian of the electron in
the magnetic field. Finally, as already mentioned, we shall
construct vector coherent states using these pairs partner
Hamiltonians.

\section{VCS for SUSY quantum models}\label{SUSY_VCS}
  In this section we outline a method for building vector coherent
states (VCS)
for supersymmetric (SUSY) quantum models. A SUSY model (see, for example, \cite{junker})
consists of two Hamiltonians,
$H^{\text b}$ and $H^{\text f}$, acting on a Hilbert space $\h$ and
factorizable in the manner,
\be
  H^{\text b} = A^\dag A,  \qquad H^{\text f} = A A^\dag .
\label{SUSY_ham1}
\en
Each Hamiltonian has a purely discrete spectrum and the two spectra coincide,
except possibly, for the lowest eigenvalue. Let us denote the normalized
eigenvectors of $H^{\text b}$ by $\phi^{\text b}_n, \; n =0,1,2,
\ldots \infty$, and those of
$H^{\text f}$ by $\phi^{\text f}_n, \; n =0,1,2, \ldots \infty$.
We shall assume the lowest eigenvalue of $H^{\text b}$ to be zero and that of
$H^{\text f}$ to coincide with the first non-zero eigenvalue of $H^{\text b}$.
Thus, we write $\varepsilon_n$, with $\varepsilon_0 = 0$, for the eigenvalues
corresponding to the eigenvectors $\phi^{\text b}_n , \;\;n=0,1,2, \ldots$,
and $\varepsilon_{n+1}$ for the eigenvalues corresponding to the eigenvectors
$\phi^{\text f}_n ,\;\; n=0,1,2, \ldots$. The operators $A$ and $A^\dag$
act on the eigenvectors in the manner,
\be
  A\phi^{\text b}_n = \sqrt{\varepsilon_n}\;\phi^{\text f}_{n-1}, \quad
  A\phi^{\text b}_n = 0, \qquad
A^\dag \phi^{\text f}_n = \sqrt{\varepsilon_{n+1}}\;\phi^{\text b}_{n+1},
\qquad n=0,1,2, \ldots ,
\label{ladder_op1}
\en
and each set of eigenvectors forms an orthonormal basis for $\h$. The
full SUSY Hamiltonian, $H^{\text{SUSY}}$ is then defined as
\be
  H^{\text{SUSY}} = \begin{pmatrix} H^{\text b} & 0 \\ 0 & H^{\text f}
  \end{pmatrix} = \begin{pmatrix} A^\dag A & 0 \\ 0 & AA^\dag
  \end{pmatrix}
\label{susy_ham} \en on the Hilbert space $\h^{\text{SUSY}} =
\mathbb C^2 \otimes \h$. The Hamiltonian  can also be written as
$H^{\text{SUSY}} =\left\{Q^\dagger,\,Q\right\}$, where
$Q=\begin{pmatrix} 0 & 0 \\ A& 0
  \end{pmatrix} $ and $Q^\dagger$ are the
{\em supercharges}. On $\h^{\text{SUSY}}$ we define the vectors \be
\bPhi^{\text b}_n = \begin{pmatrix} \phi^{\text b}_n \\
0\end{pmatrix}, \qquad \bPhi^{\text f}_n = \begin{pmatrix} 0 \\
\phi^{\text f}_n\end{pmatrix}, \qquad n=0,1,2, \ldots ,
\label{susy_basis} \en which together form an orthonormal basis for
this Hilbert space.

\subsection{Construction of the VCS}

Vector coherent states (VCS), of the type we are about to construct
here, have been introduced in \cite{{alienggaz},{thirali}} and we
shall follow the method outlined there to build  vector coherent
states for SUSY systems.  We start by defining the vectors, \be
\bPsi_0 =
\begin{pmatrix} \phi^{\text b}_0 \\ 0\end{pmatrix}, \qquad \bPsi_n =
\bPhi^{\text b}_n \oplus \bPhi^{\text f}_{n-1}
 = \begin{pmatrix} \phi^{\text b}_n \\ \phi^{\text f}_{n-1}\end{pmatrix},
 \qquad n=1,2,3, \ldots .
\label{VCS_basis}
\en
These vectors are mutually orthogonal but not all normalized:
$$ \Vert \bPsi_0\Vert^2 =1 , \qquad \Vert \bPsi_n\Vert^2 = 2, \qquad
  n =1,2,3, \ldots .
$$
However, they are eigenvectors of the SUSY Hamiltonian:
\be
  H^{\text{SUSY}}\bPsi_n = \varepsilon_n\bPsi_n , \qquad n=0,1,2,
  \ldots,
\label{ev-eqn} \en but they do not span all of $\h^{\text{SUSY}}$
since, for instance, the vector $\begin{pmatrix} 0\\ \phi^{\text
f}_0 \end{pmatrix}$ belongs to $\h^{\text{SUSY}} $ but cannot be
written as a linear combination of the $\bPsi_n$'s.

Next let $\lim_{n\rightarrow\infty}\varepsilon_n = L$, which could
be infinity, and define the domain $\mathcal D = \{ z \in \mathbb
C\mid \vert z\vert < \sqrt{L}
  \} \subseteq \mathbb C$. We also assume that the sequence
$\{\varepsilon_n !\}_{n=0}^\infty$, where, by definition
$\varepsilon_0 ! = 1$ and $\varepsilon_n ! =
\varepsilon_1\varepsilon_2\varepsilon_3\ldots \varepsilon_n ,\;
n=0,1,2, \dots$, is a moment sequence. This means that we assume
that there exists a measure $d\lambda$ on $(0,\sqrt{L})$ such that
\be
  2\pi\int_0^{\sqrt{L}} r^{2n}\; d\lambda (r) = \varepsilon_n ! , \qquad
n=0,1,2, \ldots .
\label{mom-prob1}
\en
Vector coherent states $\vert z, \overline{z}\rangle \in \h^{\text{SUSY}}$
are now defined, for each $z\in \mathcal D$, as
\bea
  \vert z, \overline{z}\rangle &=&
  \mathcal N(\vert z\vert^2)^{-\frac 12}\sum_{n=0}^\infty
  \frac {\mathfrak Z^n}{\sqrt{\varepsilon_n !}} \bPsi_n, \nonumber\\
  & = & \mathcal N(\vert z\vert^2)^{-\frac 12}\left[\sum_{n=0}^\infty
  \frac {z^n}{\sqrt{\varepsilon_n !}} \bPhi^{\text b}_n  +
  \sum_{n=0}^\infty
  \frac {\overline{z}^{n+1}}{\sqrt{\varepsilon_{n+1} !}} \bPhi^{\text f}_n
  \right] , \qquad \mathfrak Z = \begin{pmatrix} z & 0 \\ 0 & \overline{z}
\end{pmatrix},
\label{VCS1}
\ena
where the normalization constant,
\be
  \mathcal N (\vert z\vert^2) = 1 + 2 \sum_{n=1}^\infty
  \frac {\vert z\vert^{2n}}{\varepsilon_n !},
\label{normaliz} \en is chosen so that $\langle z,\overline{z}\mid
z, \overline{z}\rangle =1$, independently of $z\in\D$. Notice that
this series converges for all $z\in\mathcal D$. Defining the measure
$$ d\mu (z , \overline{z} ) = d\lambda (r) \; d\theta , \qquad \text{where}
\qquad z = re^{i\theta} , $$
it is easy to verify that these VCS satisfy the resolution of the identity,
\be
 \int_\mathcal D \vert z, \overline{z}\rangle\langle z, \overline{z}\vert\;
  \mathcal N(\vert z\vert^2)\; d\mu (z, \overline{z} ) = I_{\h^{\text{SUSY}}} ,
\label{vcs_resolid}
\en
on $\h^{\text{SUSY}}$. We shall call the vectors (\ref{VCS1})  {\em SUSY
associated VCS}\/. The term vector coherent state reflects the fact that they can also
be written as the two-component vectors:
\be
  \vert z, \overline{z}\rangle =
  \mathcal N(\vert z\vert^2)^{-\frac 12}\sum_{n=0}^\infty \; \begin{pmatrix}
   \displaystyle \frac {z^n}{\sqrt{\varepsilon_n !}}\phi_n^{\text{b}}\\[12pt]
    \displaystyle\frac {\overline{z}^{n+1}}{\sqrt{\varepsilon_{n+1} !}}\phi_n^{\text{f}}
    \end{pmatrix}\; .
\label{VCS-2comp}
\en

\subsection{Holomorphic representation}

  Let us re-emphasize that the VCS (\ref{VCS1}) are built using eigenvectors
of the SUSY Hamiltonian, with the degeneracy of the levels $\varepsilon_n,\;
n = 1,2,3, \ldots$, reflected in the choice of the vectors $\bPsi_n , \;
n = 1,2,3, \ldots$. We proceed to study some analytic features of these
VCS. Consider the Hilbert space $L^2(\mathcal D , d\mu (z, \overline{z} ))$,
in which we identify the two subspaces, $\h^{\text b}_{\text{hol}}$,
consisting of all functions analytic in $z$, including the constant
function and $\h^{\text f}_{\text{hol}}$, consisting of all functions
analytic in $\overline{z}$, excluding the constant function. Clearly, the
two subspaces are mutually orthogonal. We write $\h_{\text{hol}} =
\h^{\text b}_{\text{hol}} \oplus \h^{\text f}_{\text{hol}}$, for the subspace
consisting of all functions either analytic or anti-analytic in $z$. Let
$\mathbb P^{\text b}_{\text{hol}}$ and $\mathbb P^{\text f}_{\text{hol}}$ be
the corresponding projection operators:
\be
  \mathbb P^{\text b}_{\text{hol}}\h_{\text{hol}} = \h^{\text b}_{\text{hol}},
  \qquad
\mathbb P^{\text f}_{\text{hol}}\h_{\text{hol}} = \h^{\text f}_{\text{hol}}.
\label{proj_ops}
\en
Note that the Hilbert space $\h^{\text{SUSY}} = \mathbb C^2 \otimes \h$
can also be written as the direct sum
\be
 \h^{\text{SUSY}} = \h^{\text b}_{\text{SUSY}} \oplus
          \h^{\text f}_{\text{SUSY}},
\label{susy_decomp}
\en
of a bosonic subspace $\h^{\text b}_{\text{SUSY}}$, spanned by the vectors
$\bPhi^{\text b}_n$ and a fermionic subspace $\h^{\text f}_{\text{SUSY}}$,
spanned by the vectors $\bPhi^{\text f}_n$.

In view of the resolution of the identity (\ref{vcs_resolid}), the
mapping
\be
  W: \h^{\text{SUSY}} \longrightarrow \h_{\text{hol}}, \qquad
    (W\bPhi) (z, \overline{z}) =  \mathcal N (\vert z\vert^2)^{1/2}
    \langle \overline{z}, z \mid \bPhi \rangle,
\label{hol_isom} \en
where the order of $z$ and $\overline{z}$ is
important,  is unitary, and maps the bosonic sector $\h^{\text
b}_{\text{SUSY}}$ onto the subspace $\h^{\text b}_{\text{hol}}$ of
analytic functions in $z$ (including the constant function) and the
fermionic sector $\h^{\text f}_{\text{SUSY}}$ onto the subspace
$\h^{\text f}_{\text{hol}}$ of analytic functions in $\overline{z}$
(excluding the constant function). It is easy to see that under this
mapping the basis vectors $\bPhi^{\text b}_n$ and $\bPhi^{\text
f}_n$ transform into the monomials,
\bea
  &  & (W\bPhi^{\text b}_n)(z,\overline{z}) =
  \frac {z^n}{\sqrt{\varepsilon_n !}} := \bxi_n (z), \nonumber\\
    &
  & (W\bPhi^{\text f}_n)(z,\overline{z}) =
  \frac {\overline{z}^{n+1}}{\sqrt{\varepsilon_{n+1} !}}
  = \overline{\bxi_{n+1} (z)},
  \qquad n=0,1,2, \dots ,
\label{monomials1} \ena
 so that the vectors $\bPsi_n$, used to
construct the VCS, transform to \bea
(W\bPsi_0)(z, \overline{z})& = &\bxi_0 (z) = 1, \nonumber\\
 (W\bPsi_n)(z, \overline{z}) & = & \bxi_n (z) + \overline{\bxi_{n-1} (z)} =
 \frac {z^n + \overline{z}^n}{\sqrt{\varepsilon_n !}}, \qquad n=1,2, \dots.
\label{monomials2}
\ena
We shall then write,
\be
  \vert z, \overline{z}\rangle_\text{hol} :=
W\vert z, \overline{z}\rangle =
  \mathcal N(\vert z\vert^2)^{-\frac 12}\left[\sum_{n=0}^\infty
  \frac {z^n}{\sqrt{\varepsilon_n !}} \bxi_n +
  \sum_{n=1}^\infty
  \frac {\overline{z}^n}{\sqrt{\varepsilon_{n} !}} {\overline{\bxi}}_{n}
  \right] ,
\label{hol_susy_vcs}
\en
Also writing,
\be
  Q_\text{hol} = W\,Q\,W^{-1} , \qquad
  Q_\text{hol}^\dag = W\,Q^\dagger\,W^{-1},
\label{su-charges}
\en
for the `holomorphic supercharges', we see that they act on the vectors
$\bxi_n$ as follows:
$$ Q_\text{hol}\left(\frac {z^n}{\sqrt{\varepsilon_n !}}\right)
  = \frac {\overline{z}^n}{\sqrt{\varepsilon_{n-1}!}},
  \qquad
   Q_\text{hol}^\dag\left(\frac {\overline{z}^n}{\sqrt{\varepsilon_n !}}\right)
  = \frac {z^n}{\sqrt{\varepsilon_{n-1}!}}, \quad  n=1,2,3, \ldots ,
$$
and

$$
Q_\text{hol}\bxi_0(z) = Q_\text{hol}^\dag\bxi_0(z)
= Q_\text{hol}\bxi_n(\overline{z}) = Q_\text{hol}^\dag\bxi_n(z) = 0,
\qquad n=1,2, 3,\ldots ,
$$
so that, apart from the constant function, they basically
interchange the holomorphic and antiholomorphic sectors. Clearly,
\be \{Q_\text{hol}^\dag, Q_\text{hol}\} = Q_\text{hol}^\dag
Q_\text{hol} + Q_\text{hol} Q_\text{hol}^\dag = WH^{\text
{SUSY}}W^{-1} =:
  H^{\text {SUSY}}_{\text {hol}} .
\label{susy_ham2}
\en
Note that the ground state wave function, $\bxi_0$, used in constructing
the VCS in (\ref{hol_susy_vcs}), satisfies
$$ Q_\text{hol}\bxi_0 = Q_\text{hol}^\dag\bxi_0 = 0, $$
which is reflective of the fact that we are using a model where SUSY is
unbroken.

\subsection{Creation and annihilation operators}
Suppose we define the formal {\em annihilation operator}, $\mathcal A$,
by its action on the VCS (\ref{VCS1}),
\be
\mathcal A\vert z , \overline{z}\rangle =
  \mathfrak Z \vert z , \overline{z}\rangle ,
\label{susy_annih_op1} \en where, on the right hand side,
multiplication of the vector $\vert z , \overline{z}\rangle$,
considered as an element in  $\mathbb C^2$, by the matrix $\mathfrak
Z$ is implied. It is easily seen that the above equation is
recovered by the following action of $\mathcal A$ on the vectors
$\bPsi_n$: \be \mathcal A \bPsi_0 =0,\quad \mathcal A \bPsi_n =
\sqrt{\varepsilon_n}\bPsi_{n-1}, \qquad n=1, 2,
 \ldots ,
\label{susy_annih_op2} \en
which has the familiar form of shift
operators. We would like to define an adjoint operator, $\mathcal
A^\dag$, such that $\mathcal A^\dag\mathcal A$ would coincide with
$H^\text{SUSY}$. However, since the vectors $\bPsi_n$ do not span
the whole of $\h^{\text{SUSY}}$ and since they are not all
normalized, the usual relations, $\mathcal A^\dag \bPsi_n =
\sqrt{\varepsilon_{n+1}}\bPsi_{n+1}$, will not define the adjoint.
In fact, if we compute the adjoint of $\mathcal A$ on the subspace
generated by the orthonormal set of vectors $\bPsi_0, \frac
1{\sqrt{2}}\bPsi_n, n=1,2,3, \ldots$, we easily obtain,
$$ \mathcal A^\dag \bPsi_0 =
\sqrt{\frac {\varepsilon_1}2}\frac {\bPsi_1}{\sqrt{2}} , \qquad
\mathcal A^\dag \frac {\bPsi_n}{\sqrt{2}} =
\sqrt{\varepsilon_{n+1}}\frac {\bPsi_{n+1}}{\sqrt{2}}, \qquad
n=1,2,3, \ldots .$$ Also, it is easily checked that $\mathcal A^\dag
\mathcal A\bPsi_n = \varepsilon_n\bPsi_n, \; n=0,2,3, \ldots$, but
$\mathcal A^\dag \mathcal A\bPsi_1 = \displaystyle\frac
{\varepsilon_1}2\bPsi_1$, so that $\mathcal A^\dag \mathcal A$ does
coincide with $H^{\text{SUSY}}$ on this subspace.

 To proceed further, we first extend $\mathcal A$ and $\mathcal A^\dag$
to the entire set of basis vectors $\bPhi_n^\text{b},
\bPhi_n^\text{f}, \; n=0,1,2, \ldots , $ (see (\ref{susy_basis})),
spanning $\h^{\text{SUSY}}$. Let $a_\text{b}, a_\text{b}^\dag$
denote the usual shift operators in $\h$, acting on the normalized
eigenvectors, $\phi_n^\text{b}, \; n=0,1,2, \dots ,$ of the bosonic
Hamiltonian $H^\text{b} = A^\dag A$ (see
(\ref{SUSY_ham1})-(\ref{ladder_op1})): \be
 a_\text{b}\phi_0^\text{b} =0,\quad a_\text{b}\phi_n^\text{b} =
 \sqrt{\varepsilon_n}\phi_{n-1}^\text{b}, \,n=1,2,3,\ldots,
\quad
  a_\text{b}^\dag\phi_n^\text{b} = \sqrt{\varepsilon_{n+1}}\phi_{n+1}^\text{b},
\qquad n=0,1,2, \ldots . \label{bosonic_shifts1}
\en
Then
$a_\text{b}^\dag a_\text{b} = A^\dag A = H^\text{b}$. We now want to
define similar operators $a_\text{f}, a_\text{f}^\dag$, acting on
the normalized eigenvectors $\phi_n^\text{f}, \; n=0,1,2, \ldots,$
of the fermionic Hamiltonian $H^\text{f} = AA^\dag$, such that
$a_\text{f}^\dag a_\text{f} = AA^\dag$. Note, however, that the
lowest eigenvalue of $H^\text{f}$ is $\varepsilon_1 \neq 0$.

  Let us start by defining
\bea
a_\text{f}\phi_n^\text{f} & = &
\sqrt{\varepsilon_{n+1}}\phi_{n-1}^\text{f},
\quad n=1,2,3, \ldots , \qquad a_\text{f}\phi_0^\text{f} = 0,\nonumber\\
a_\text{f}^\dag\phi_n^\text{f} & = &
\sqrt{\varepsilon_{n+2}}\phi_{n+1}^\text{f}, \quad n=0,1,2, \ldots .
\label{fermionic_shifts1}
\ena
However, this gives $a_\text{f}^\dag
a_\text{f}\phi_0^\text{f} = 0$ and $a_\text{f}^\dag
a_\text{f}\phi_n^\text{f} = \varepsilon_n\phi_n^\text{f}, \;
n=1,2,3, \ldots$. In order to correct for the appearance of $0$ and
not $\varepsilon_1$ as the lowest eigenvalue, it is
convenient to extend the operators $a_\text{f}^\dag,
a_\text{f}$ to a larger Hilbert space. To do this, we adjoin an
abstract vector $\chi$ to the Hilbert space $\h$ and extend its
scalar product so that $\chi$ has unit norm and is orthogonal to
$\h$ in this product. Let $\widetilde{\h}$ and
$\langle\cdot\mid\cdot\rangle^\sim $ denote this extended space and
scalar product, respectively, so that,
\be
 \langle \chi\mid\chi\rangle^\sim = 1, \qquad
\langle \chi\mid\phi\rangle^\sim = 0, \qquad \langle
\psi\mid\phi\rangle^\sim = \langle \psi\mid\phi\rangle_\h, \qquad
\forall \psi, \phi \in \h .
\label{ex-scal_prod}
\en
An arbitrary vector
$\widetilde{\phi}\in \widetilde{\h}$ has the form $\widetilde{\phi}
= u\chi + v\phi$, for some $u,v \in \mathbb C$ and $\phi \in \h$. On
$\widetilde{\h}$ we define the operators $\widetilde{a}_\text{f},
\widetilde{a}_\text{f}^\dag$ as \bea \widetilde{a}_\text{f}\chi = 0,
&\quad& \widetilde{a}_\text{f}\phi_0^\text{f} =
\sqrt{\varepsilon_1}\chi, \qquad
\widetilde{a}_\text{f}\phi_n^\text{f}=
\sqrt{\varepsilon_{n+1}}\phi_{n-1}^\text{f},\quad n= 1,2,3,
\ldots , \nonumber\\
\widetilde{a}_\text{f}^\dag\chi =
\sqrt{\varepsilon_1}\phi_0^\text{f}, & \quad &
\widetilde{a}_\text{f}^\dag\phi_n^\text{f} =
\sqrt{\varepsilon_{n+2}}\phi_{n+1}^\text{f}, \quad n=0,1,2, \ldots .
\label{fermionic_shifts2} \ena Clearly,  on $\h$ we
have
\be
 a_\text{f} = \mathbb P_\h \widetilde{a}_\text{f}\mathbb P_\h , \qquad
a_\text{f}^\dag = \mathbb P_\h \widetilde{a}_\text{f}^\dag\mathbb P_\h
  = \widetilde{a}_\text{f}^\dag\mathbb P_\h , \qquad
\widetilde{a}_\text{f}^\dag\widetilde{a}_\text{f}\mathbb P_\h =
AA^\dag , \label{restrict_ops}
\en
$\mathbb P_\h$ being the
projector from $\widetilde{\h}$ to $\h$ which acts as
$\mathbb
P_\h\widetilde\Phi=\widetilde\Phi-<\chi,\widetilde\Phi>\chi$. We
similarly extend the fermionic subspace $\h^\text{f}_\text{SUSY}$ of
$\h^\text{SUSY}$ (see (\ref{susy_decomp})), by adding the vector \be
  \bPhi_{00} = \begin{pmatrix} 0 \\ \chi\end{pmatrix},
\label{extra_vect}
\en
and extending the scalar product, as before, so that $\Phi_{00}$ has
unit norm and is orthogonal to $\h^\text{f}_\text{SUSY}$. We denote the
extended space by $\widetilde{\h}^\text{f}_\text{SUSY}$ and write
$$
  \widetilde{\h}^\text{SUSY} = \h^\text{b}_\text{SUSY} \oplus
   \widetilde{\h}^\text{f}_\text{SUSY}. $$

   On this extended space $\widetilde{\h}^\text{SUSY}$, we now define
the two two operators, \be
\widetilde{\mathcal A} = \begin{pmatrix} a_\text{b} & 0 \\
             0 & \widetilde{a}_\text{f} \end{pmatrix}, \qquad
\widetilde{\mathcal A}^\dag = \begin{pmatrix} a_\text{b}^\dag & 0 \\
             0 & \widetilde{a}_\text{f}^\dag \end{pmatrix} ,
\label{susy_shifts1} \en so that, denoting the projector from
$\widetilde{\h}^\text{SUSY}$ to $\h^\text{SUSY}$ by
$\widetilde{\mathbb P}$, we set \be \mathcal A_\text{SUSY} =
\widetilde{\mathbb P}\,\widetilde{\mathcal A}\,\widetilde{\mathbb
P}, \qquad \mathcal A^\dag_\text{SUSY} = \widetilde{\mathbb
P}\,\widetilde{\mathcal A}^\dag\,\widetilde{\mathbb P}.
\label{susy_shifts2} \en Clearly, $\mathcal A_\text{SUSY}$ is the
extension to $\widetilde\h^\text{SUSY}$  of the operator $\mathcal A$ defined in
 (\ref{susy_annih_op1})-(\ref{susy_annih_op2}). Also, these operators
 act on the vectors $\bPhi_n^\text{b}, \bPhi_n^\text{f}, \; n=0, 1,2,3,
 \ldots , $ in the expected manner:
\begin{align}
  & \mathcal A_\text{SUSY}\bPhi_0^\text{b} =0,
  \quad\mathcal A_\text{SUSY}\bPhi_n^\text{b} = \sqrt{\varepsilon_n}
  \;\bPhi_{n-1}^\text{b},\quad n=1,2,\ldots,\nonumber\\
& \mathcal A_\text{SUSY}^\dag\bPhi_n^\text{b}  =
\sqrt{\varepsilon_{n+1}}\;
     \bPhi_{n+1}^\text{b},\quad n=1,2, \ldots  \nonumber\\
& \mathcal A_\text{SUSY}\bPhi_0^\text{f} = 0,\quad\mathcal
A_\text{SUSY}\bPhi_n^\text{f} = \sqrt{\varepsilon_{n+1}}\;
\bPhi_{n-1}^\text{f}, \quad n=1,2,3, \ldots , \nonumber\\
& \mathcal A_\text{SUSY}^\dag\bPhi_n^\text{f}  = \sqrt{\varepsilon_{n+2}}\;
     \bPhi_{n+1}^\text{f},\quad n=0,1,2, \ldots .
\label{susy_shifts3}
\end{align}
The SUSY Hamiltonian can now be written as (see (\ref{susy_ham})):
\be H^\text{SUSY} = \widetilde{\mathbb P}\begin{pmatrix}
  a_\text{b}^\dag a_\text{b} & 0\\ 0 &
  \widetilde{a}_\text{f}^\dag \widetilde{a}_\text{f} \end{pmatrix}
  \widetilde{\mathbb P} = \widetilde{\mathbb P}\widetilde{\mathcal A}^\dag
  \widetilde{\mathcal A}\widetilde{\mathbb P} .
\label{susyham3}
\en
Note that while this Hamiltonian now appears in the form $B^\dag B$, with
$B= \widetilde{\mathcal A}\widetilde{\mathbb P}$, the range of the operator
$B$ includes the additional vector $\bPhi_{00}$ and the domain of $B^\dag$
is the extended space $\widetilde{\h}^\text{SUSY}$.

\subsection{VCS on the extended space}

  It is interesting to  define now VCS on the enlarged Hilbert space
$\widetilde{\h}^\text{SUSY}$, which extend the SUSY associated VCS
introduced in (\ref{VCS1}). We define the vectors (see
(\ref{VCS_basis})),
$$
  \widetilde{\bPsi}_0 = \frac{1}{\sqrt{2}}\, \begin{pmatrix} \phi_0^\text{b} \\ \chi \end{pmatrix},
  \qquad \widetilde{\bPsi}_n = \frac{1}{\sqrt{2}}\,\bPsi_n, \quad n=1,2,3, \ldots ,
  $$
 and set
\be
\vert z, \overline{z}\rangle\!^\sim = \mathcal N (\vert z\vert^2
)^{-\frac 12} \sum_{n=0}^\infty \frac {\mathfrak
Z^2}{\varepsilon_n!} \widetilde{\bPsi}_n, \qquad \mathfrak Z =
\begin{pmatrix} z & 0 \\ 0 & \overline{z} \end{pmatrix} ,
\label{ext_VCS} \en with $\mathcal N$ defined as before (see
(\ref{normaliz})). These vectors are normalized. Indeed,
$$ ^\sim\!\langle z, \overline{z}\mid z, \overline{z} \rangle\!^\sim = 1 , $$
and we still have a resolution of the identity, \be
  \int_\mathcal D \vert z, \overline{z} \rangle\!^\sim\!\;^\sim\!\langle z,
  \overline{z}\vert\;\mathcal N (\vert z\vert^2 )\;
    d\mu (z, \overline{z}) = I_{\widetilde{\h}^\text{SUSY}} ,
\label{ext_VCS2}
\en
on the enlarged space $\widetilde{\h}^\text{SUSY}$. The physical
or SUSY associated VCS (\ref{VCS1}) are now obtained by simple projection,
\be
\vert z, \overline{z}\rangle =  \widetilde{\mathbb P}
   \vert z, \overline{z}\rangle\!^\sim, \qquad z \in \mathcal D .
\label{projec_VCS}
\en
Furthermore, we easily verify the relations,
$$
\widetilde{\mathcal A}\vert z, \overline{z}\rangle\!^\sim = \mathfrak Z
\vert z, \overline{z}\rangle\!^\sim, \quad
\widetilde{\mathcal A}\widetilde{\bPsi}_n =
\sqrt{\varepsilon_n}\widetilde{\bPsi}_{n-1}, \quad
\widetilde{\mathcal A}^\dag\widetilde{\bPsi}_n =
\sqrt{\varepsilon_{n+1}}\widetilde{\bPsi}_{n+1} .$$

  Finally let us note that the appearance of the vectors $\chi$ and
$\bPhi_{00}$ in the discussion (see (\ref{extra_vect})) above is not
entirely spurious. Indeed, the existence of such a vector is guaranteed
when SUSY is not broken. In a  generic SUSY model, the two operators,
$A$ and $A^\dag$ act on the Hilbert space $\h = L^2 (\mathbb R, dx)$ and
have the form:
\be
 A = \frac \hbar{\sqrt{2m}}\frac d{dx} + W(x), \qquad
 A^\dag = -\frac \hbar{\sqrt{2m}}\frac d{dx} + W(x),
\label{generic_susy_shifts} \en where $W(x)$ is a real
`superpotential'. Since we are assuming that the bosonic ground
state $\phi_0^\text{b}$ is an eigenstate of $H^\text{b} = A^\dag A$
with eigenvalue $\varepsilon_0 =0$, this wave function satisfies
$$ A\phi_0^\text{b} =\frac \hbar{\sqrt{2m}}\frac d{dx}\phi_0^\text{b}
    + W(x)\phi_0^\text{b} = 0, $$
from which we get \be
  \phi_0^\text{b}(x) = \exp\left[-\frac {\sqrt{2m}}\hbar \int_0^x
         W(x')\; dx' \right].
\label{expl_bos_gr_st1}
\en
Next, if we try to find a vector $\chi$ which would correspond to the zero
eigenvalue of $AA^\dag$, we need to solve
$$
 A^\dag\chi =-\frac \hbar{\sqrt{2m}}\frac d{dx}\chi
    + W(x)\chi = 0. $$
We thus find \be
  \chi(x) = \exp\left[\frac {\sqrt{2m}}\hbar \int_0^x
         W(x')\; dx' \right],
\label{expl_bos_gr_st2}
\en
which will generally not be square-integrable, if the solution in
(\ref{expl_bos_gr_st1}) is square-integrable. It is this vector that we adjoined to the
Hilbert space $\h$ to obtain the space $\widetilde{\h}$ above, but of
course, we had to extend the scalar product of $\h = L^2 (\mathbb R, dx)$
to accomodate it (see (\ref{ex-scal_prod})). Thus, the extended VCS in (\ref{ext_VCS})
include this ``unphysical'' vector which is not $L^2$-normalizable.

\subsection{An alternative realization}

Before ending this discussion on the  general construction of SUSY associated VCS, let us
note that the vectors (\ref{hol_susy_vcs}) can also be written in the  standard SUSY forrmalism,
using anticommuting variables. We start by introducing the complex Grassmann variables $\zeta,
\overline{\zeta}$ which satisfy
\be
 \zeta^2 = \overline{\zeta}^2 = 0 , \qquad \zeta\overline{\zeta} = -\overline{\zeta}\zeta \; ,
\label{grassmann1}
\en
and with respect to the formal measure $d\zeta$ have the ``fermionic (Berezin) integration'' properties:
\be
  \int_{\mathbb C^{1\mid 1}} \zeta\; d\zeta =  \int_{\mathbb C^{1\mid 1}}
 \overline{\zeta}\; d\zeta =  \int_{\mathbb C^{1\mid 1}} d\zeta = 0 , \qquad
   \int_{\mathbb C^{1\mid 1}} \overline{\zeta}\zeta\; d\zeta = 1 \; ,
\label{grassmann2}
\en
$\mathbb C^{1\mid 1}$ denoting the formal domain of the Grassmann variable $\zeta$.
We consider now the Hilbert space $\h^\text{b}_\text{hol}$ of holomorphic functions,
defined earlier,
and its subspace $\h^1_\text{hol}$ which consists of all functions in $\h^\text{b}_\text{hol}$ except for
the constant function. Consider next functions in the two variables $z, \zeta$, of the type $\bxi (z, \zeta)
= \bxi^\text{b} (z) + \zeta\bpsi (z)$, with $\bxi^\text{b} \in \h^\text{b}_\text{hol}$
and $\bpsi \in \h^1_\text{hol}$. These
functions form  a Hilbert space with respect to the scalar product
\bea
\langle \bxi_1 \mid \bxi_2\rangle & = &  \int_{\mathcal D^{1\mid 1}} \overline{\bxi_1 (z, \zeta)}\; \bxi_2 (z, \zeta)\;
[1 + \overline{\zeta}\zeta]\;d\zeta\;d\mu (z, \overline{z})\nonumber\\
& = & \int_{\mathcal D}\overline{\bxi^\text{b}_1(z)}\;
\bxi^\text{b}_2 (z)\; d\mu (z, \overline{z}) + \int_{\mathcal D}\overline{\bpsi_1(z)}\;
\bpsi_2 (z)\; d\mu (z, \overline{z})\; ,
\label{grassmann-sc-prod}
\ena
where $\mathcal D^{1\mid 1}$ now denotes the joint domain of the variables $\zeta$ and
$z$.
We denote this Hilbert space by $\mathfrak K^\text{SUSY}$ and note that
(\ref{grassmann-sc-prod}) implies the formal orthogonal decomposition,
$\mathfrak K^\text{SUSY} \simeq \h^\text{b}_\text{hol} \oplus \h^1_\text{hol}$. The
coherent states (\ref{hol_susy_vcs}), expressed in this alternative notation now appear as
\bea
\vert z, \zeta\rangle & = & \mathcal N(\vert z\vert^2)^{-\frac 12}\left[\sum_{n=0}^\infty
  \frac {z^n}{\sqrt{\varepsilon_n !}} \bxi_n +
  \zeta\sum_{n=1}^\infty
  \frac {z^n}{\sqrt{\varepsilon_{n} !}} \bxi_{n}
  \right]\nonumber\\
  & = & \mathcal N(\vert z\vert^2)^{-\frac 12}\left[\bxi_0 + (1 + \zeta)\sum_{n=1}^\infty
  \frac {z^n}{\sqrt{\varepsilon_n !}} \bxi_n \right] ,
\label{susy-hol-vcs}
\ena
and they satisfy the formal resolution of the identity,
\be
  \int_{\mathcal D^{1\mid 1}}\vert z, \zeta \rangle\langle z, \zeta \vert \;
  \mathcal N(\vert z\vert^2)\; [\overline{\zeta}\zeta - 1 ]
  \; d\zeta\; d\mu (\overline{z}, z ) = I_{\h^\text{b}_\text{hol}}
    \oplus I_{\h^1_\text{hol}} \simeq I_{\mathfrak K^\text{SUSY}}\; ,
\label{susy-hol-resolid}
\en
which is to be compared to (\ref{vcs_resolid}).

\section{Landau levels}

We proceed to apply the theory of supersymmetric coherent states just developed,
to certain concrete physical models related to the quantum Hall effect and some
of its generalizations.

\subsection{Standard Landau levels}

The Hamiltonian of a single electron, moving on a two-dimensional plane and
subject to a uniform magnetic field along the $z$-direction, is given by
\begin{equation}
H_0={\frac 12}\,\left(\underline p+\underline A(r)\right)^2={\frac 12}\;
\left(p_x-{\frac y2}\right)^2+{\frac 12}\,\left(p_y+{\frac x2}
\right)^2, \label{21}
\end{equation}
where we have used minimal coupling and the symmetric gauge
$\vec A=\frac{1}{2}(-y,x,0)$.

The spectrum of this hamiltonian is easily obtained by first
introducing the new variables
  \be
\label{22}
  P'= p_x-y/2, \hspace{5mm}     Q'= p_y+x/2.
  \en
In terms of $P'$ and $Q'$ the single electron hamiltonian, $H_0$,
can be rewritten as
 \be
\label{23}
  H_{0}=\frac{1}{2}(Q'^2 + P'^2).
  \en
The transformation (\ref{22}) is part of a canonical map from the
phase space variables $(x,y,p_x,p_y)$ to $(Q,P,Q',P')$, where
 \be
\label{24}
   P= p_y-x/2, \hspace{5mm}
  Q= p_x+y/2.
   \en
Indeed, we easily see that
 $$ \begin{pmatrix} Q\\Q'\\P\\P' \end{pmatrix}   = S\begin{pmatrix} x\\y\\p_x\\p_y \end{pmatrix},
 \quad \text{where} \quad S = \begin{pmatrix} 0 & \frac 12 & 1 & 0 \\ \frac 12 & 0& 0 &1
  \\ -\frac 12 & 0& 0 &1 \\ 0 & -\frac 12 & 1 & 0 \end{pmatrix},
 $$
 and $S$ is a symplectic matrix:
 $$ SJS^T = J, \quad \text{with} \quad J = \begin{pmatrix} \mathbf 0 & \mathbb I_2
 \\ -\mathbb I_2 & \mathbf 0\end{pmatrix}, \quad \mathbb I_2 =
 \begin{pmatrix} 1& 0 \\ 0 & 1\end{pmatrix}.$$
Moreover, at the classical level one also verifies the invariance of the associated {\em two-form}:
$$ dx\wedge dp_x + dy\wedge dp_y = dQ\wedge dP + dQ' \wedge dP' $$
under this transformation.

  The corresponding quantized  operators  satisfy the commutation relations:
$$ [x, p_x] = [y, p_y] = i, \quad [x,p_y] = [y,p_x] = [x,y] = [p_x , p_y ] = 0, $$
and
  \be
\label{25}
 [Q,P] = [Q',P']=i, \quad  [Q,P']=[Q',P]=[Q,Q']=[P,P']=0.
  \en

As discussed extensively in the literature (see, for example, \cite{b6} and references
therein),  a wave function in the $(x,y)$-space is related to its  $PP'$-counterpart
by the formula
  \be
\label{26}
  \Psi(x,y)=\frac{e^{ixy/2}}{2\pi}\int_{-\infty}^{\infty}\,
  \int_{-\infty}^{\infty}e^{i(xP'+yP+PP')}\Psi(P,P')\,dP dP',
  \en
which can be easily inverted:
  \be
\label{27}
  \Psi(P,P')=\frac{e^{-iPP'}}{2\pi}\int_{-\infty}^{\infty}\,
  \int_{-\infty}^{\infty}e^{-i(xP'+yP+xy/2)}\Psi(x,y)\,dx dy.
  \en
  The usefulness of the $PP'$-representation has been widely analyzed
in several papers over the years,
in particular in connection with the problem of finding the ground
state for the fractional quantum Hall effect (QHE), using  techniques of
multi-resolution analysis (see \cite{b4,b5,b7,bag2005} and references therein).

It is clear that, introducing the ladder operators $B, B^\dag$ as
follows \be B=\frac{Q'+iP'}{\sqrt{2}},
\hspace{1cm}B^\dagger=\frac{Q'-iP'}{\sqrt{2}}\,\,\Rightarrow\,\,
[B,B^\dagger]= I, \label{28}\en and the hamiltonian can be written
as $H_0=B^\dagger B+\frac{1}{2}$. It is well known that for the
standard harmonic oscillator there is not much to be gained by
introducing the supersymmetric partner Hamiltonians $H^\text{b}$ and
$H^\text{f}$: indeed they are simply the same hamiltonian apart from
an additive constant. If we define $H^\text{b}=H_0-\frac{1}{2}I =
B^\dagger B$ and $H^\text{f}=H_0+\frac{1}{2}I= B B^\dagger$ then the
eigenvalues of $H^\text{b}$ are $E_n^{(\text{b})}=n$, $n\in
\mathbb{N}\cup\{0\}$, and its eigenstates are
$\Psi_n^{(\text{b})}=\frac{(B^\dagger)^n}{\sqrt{n!}}\Psi_0^{(\text{b})}$,
where $B\Psi_0^{(\text{b})}=0$, $H^\text{b}\Psi_n^{(\text{b})}=
E_n^{(\text{b})}\Psi_n^{(\text{b})}$, while for $H^\text{f}$ we have
$E_n^{(\text{f})}=E_{n+1}^{(\text{b})}=n+1$,
$n\in\mathbb{N}\cup\{0\}$, and
$\Psi_n^{(\text{f})}=\frac{1}{\sqrt{E_n^{(\text{f})}}}\,B\,\Psi_{n+1}^{(\text{b})}=
\Psi_n^{(\text{b})}$.

This illustrates what we can call {\em the triviality of the SUSY
approach} for the Hamiltonian of the standard Landau levels:
$H^\text{b}$ and $H^\text{f}$ are essentially the same operator, and
they are both very closely related to the original quantum
mechanical hamiltonian, $H_0$. Nevertheless, the formalism of
one-dimensional supersymmetry has been employed in the study of
Landau levels in a recent paper, \cite{metz}. This was done in a
rather complicated way, {\em viz} by defining a family of radial
Hamiltonians, depending on the orbital angular momentum eigenvalue
$\ell$ of the original two-dimensional system. In this way a family
of $\ell$-dependent supersymmetric partner hamiltonians were
constructed. In other words, a two-dimensional physical system was
mapped into an infinite family of one-dimensional systems.

In this paper we adopt a different point of view, using a truly
two-dimensional SUSY \cite{daspern}, which we slightly adapt to our
purposes.

It is clear that, because of the commutation rules (\ref{25}), each
Landau level is infinitely degenerate  (see, for example,
\cite{bag2005}). It is instructive to construct the vector coherent
states associated to this system, since this will also serve as a
model for the other cases, discussed below.

   Since the energy levels of $H_0$ are infinitely degenerate, we denote the
corresponding normalized eigenstates by $\mid n , k \rangle,\;  n, k
= 0, 1, 2, 3, \ldots , \infty$, with $H_0 \mid n , k \rangle = (n +
\displaystyle \frac 12)\mid n , k \rangle$ and $k$ denoting the
degeneracy parameter. These vectors form an orthonormal basis for
the Hilbert space $\h$ of the system.  Vector coherent states, for
the SUSY pair of Hamiltonians $H^\text{b}, H^\text{f}$ are now
defined in $\mathbb C^2 \otimes \h$ for each degeneracy level $k$,
following  (\ref{VCS1}), as \be
  \vert z, \overline{z}\; ; k \rangle = \mathcal N(\vert z\vert^2)^{-\frac 12}
     \sum_{n=0}^\infty
  \begin{pmatrix} \displaystyle\frac {z^n}{\sqrt{n !}} \\[8pt]
  \displaystyle\frac {\overline{z}^{n+1}}{\sqrt{(n +1)!}}\end{pmatrix}
 \otimes  \vert n, k\rangle \; , \quad
k =0, 1, 2, \ldots , \infty\; .
\label{FQHE-VCS1}
\en
Here $N(\vert z\vert^2) = 2e^{\vert z \vert^2} -1$. These vectors then satisfy the resolution
of the identity,
\be
  \sum_{k=0}^\infty \int_{\mathbb C}\vert z , \overline{z}\; ; k \rangle
\langle z, \overline{z}\; ; k\vert\;
\mathcal N(\vert z\vert^2)\; e^{-\vert z \vert^2}\; \displaystyle\frac {dx\; dy}\pi  =
\begin{pmatrix} I_\h & 0 \\ 0 & I_\h \end{pmatrix}\;,
\quad z = x + i y\;
\label{FQHE-reloid1}
\en

\subsection{Generalized Landau levels}

This section is devoted to the analysis of some quantum mechanical
models naturally arising from $H_0$ when SUSY is taken into account.

   Introducing the function $\vec
W_0=-\frac{1}{2}(x,y,0)=(W_{0,1},W_{0,2},0)$ we may rewrite the
operators in (\ref{22}) and (\ref{24}) as \be
P'=p_x+W_{0,2},\hspace{4mm}Q'=p_y-W_{0,1},\hspace{4mm}
P=p_y+W_{0,1},\hspace{4mm}Q=p_x-W_{0,2}.\label{31}\en This
definition can be extended as follows \be
p'=p_x+W_{2},\hspace{4mm}q'=p_y-W_{1},\hspace{4mm}
p=p_y+W_{1},\hspace{4mm}q=p_x-W_{2},\label{32}\en introducing a
vector superpotential $\vec W=(W_{1},W_{2},0)$. Our notation is the
following: small letters (like $q, p, q'$ and
$p'$) refer to a generic superpotential $\vec W$, while
capital  letters (like $Q, P, Q'$ and $P'$) refer
to the particular choice of superpotential $\vec W_0$, i.e. when we
consider the standard Landau levels.

We now put
\be e=-\frac{1}{\sqrt{2}}(q'+ip'),  \hspace{3mm}
e^\dagger=-\frac{1}{\sqrt{2}}(q'-ip'),\hspace{3mm}
k=-\frac{1}{\sqrt{2}}(q+ip),  \hspace{3mm}
k^\dagger=-\frac{1}{\sqrt{2}}(q-ip), \label{33}\en where the overall
minus sign has been introduced everywhere in order to preserve the same
notation as in \cite{daspern}. Thus,
$E=-\frac{1}{\sqrt{2}}(Q'+iP')=-B,$
$E^\dagger=-\frac{1}{\sqrt{2}}(Q'-iP')=-B^\dagger,$
$K=-\frac{1}{\sqrt{2}}(Q+iP),$ and
$K^\dagger=-\frac{1}{\sqrt{2}}(Q-iP)$. The following commutation
rules can be easily obtained: \be \left\{
\begin{array}{ll}
[q,p]=[q',p']=-i\vec\nabla\cdot\vec W, \\
\,\![p',p]=[q',q]=-i(\partial_xW_1)+i(\partial_yW_2),\\
\,\![q',p]=-2i(\partial_yW_1), \hspace{4mm}
[p',q]=2i(\partial_xW_2),\label{34}
\end{array}
\right.
 \en
which immediately imply
 \be \left\{
\begin{array}{ll}
[e,e^\dagger]=[k,k^\dagger]=-\vec\nabla\cdot\vec W, \\
\,\![k,e]=(\partial_xW_2)-(\partial_yW_1),\\
\,\![k,e^\dagger]=-(\partial_xW_2)-(\partial_yW_1).
\end{array}\label{35}
\right.
 \en
It is easy to check that if we take $\vec W=\vec W_0$, these
commutation relations yield those of the previous subsection.  Note
also, that classically the transformation (\ref{32}) is canonical,
i.e., $dx\wedge dp_x + dy\wedge dp_y = dQ\wedge dP + dQ'\wedge dP'$,
if and only if $\vec{W} = \vec{W}_0$, so that $\vec{\nabla}\cdot
\vec{W} = -1$.

We now introduce two pairs of supersymmetric partner Hamiltonians
\be h^\text{b}=e^\dagger e, \quad h^\text{f}=ee^\dagger,\,\hspace{2cm} {\mathfrak
h}^\text{b}=k^\dagger k,\quad {\mathfrak h}^\text{f}=k\,k^\dagger,\label{36}\en
which are related to each other by
 \be h^\text{b}-h^\text{f}={\mathfrak
h}^\text{b} -{\mathfrak h}^\text{f}=-\vec\nabla\cdot\vec W\label{37}\en
Let us  focus our attention on $h^\text{b}$ and  $h^\text{f}$ which can also be
written as
\be\left\{
\begin{array}{ll}
h^\text{b}=e^\dagger\,e=\frac{1}{2}(p_x+W_2)^2+\frac{1}{2}(p_y-W_1)^2+
\frac{1}{2}\vec\nabla\cdot\vec W,\\
h^\text{f}=e\,e^\dagger=\frac{1}{2}(p_x+W_2)^2+\frac{1}{2}(p_y-W_1)^2-
\frac{1}{2}\vec\nabla\cdot\vec W \; .
\end{array}\label{38}
\right.
 \en
The {\em capital counterparts} of these relations turn out to be
$H^\text{b}=E^\dagger\,E = \frac{1}{2}(p_x-y/2)^2+
\frac{1}{2}(p_y+x/2)^2-\frac{1}{2}\, I =H_0-\frac{1}{2}\,I$ and
$H^\text{f}=E^\dagger\,E=\frac{1}{2}(p_x-y/2)^2+
\frac{1}{2}(p_y+x/2)^2+\frac{1}{2}\, I =H_0+\frac{1}{2}\,I$, which
we have already discussed. The analysis of ${\mathfrak h}^\text{b}$
and ${\mathfrak h}^\text{f}$ is not significantly different from
that of $h^\text{b}$ and $h^\text{f}$, and will be omitted here.

If we now compare the expression of $H_0$ in (\ref{21}) with those
of $h^\text{b} -\frac{1}{2}\vec\nabla\cdot\vec W$ and
$h^\text{f}+\frac{1}{2}\vec\nabla\cdot\vec W$ in (\ref{38}), it is
easy to see that the superpotential $\vec W$ is related to the
vector potential and, therefore, to the magnetic field, as follows:
\be A_1=W_2,\,A_2=-W_1, \Rightarrow \vec B=\vec\nabla\wedge\vec
A=-\hat k(\vec\nabla\cdot\vec W),\label{39}\en where $\hat
k=(0,0,1)$. Needless to say that, when $\vec W=\vec W_0$,  the
situation reverts to the one discussed in the previous section.
However, for different choices of $\vec W$, the supersymmetry
produces inequivalent conjugate Hamiltonians which, in some sense,
extend the original operator $H_0$. Our goal is to find explicit
examples of such partner Hamiltonians, whose spectra are completely
discrete, with each energy level being infinitely degenerate,
and which therefore come under the purview of both a
generalized quantum Hall effect and a proper supersymmetric theory.

\subsubsection{Case 1: $\vec\nabla\cdot\vec W=0$.}

At  first sight this choice may seem rather trivial since, because
of (\ref{39}), it corresponds to a zero magnetic field: $\vec B=\vec
0$. However, as we show below, some non trivial mathematics and
physics do nevertheless appear.

Since $\vec\nabla\cdot\vec W=0$ we have: \be \vec B=\vec 0, \quad
h^\text{b} = h^\text{f}
=\frac{1}{2}(p_x+W_2)^2+\frac{1}{2}(p_y-W_1)^2,\quad
[e,e^\dagger]=[k,k^\dagger]=0,\label{310} \en while, on the other
hand, $[k,e]$ and $[k,e^\dagger]$ need not to be zero. To be
concrete, let us fix $\vec W=\frac{1}{2}(-y,x,0)$, as an example.
With this choice we have that $h^\text{b} =h^\text{f}
=\frac{1}{2}(p_x+x/2)^2+\frac{1}{2}(p_y+y/2)^2 , \quad
[e,e^\dagger]=[k,k^\dagger]=[k,e^\dagger]=0$ while $[k,e]= I $.

If we now introduce the  vectors $\varphi_0^{(k)}$ and
$\varphi_0^{(e)}$, such that $k\varphi_0^{(k)}= e\varphi_0^{(e)}=0$,
and the two operators, $X_+=ke, \;\; X_-=ek$, we see that:
\begin{itemize}
\item these two operators are related to each other: $X_+-X_-= I$;
\item if the vectors $\varphi_n^{(e)}=k^n\varphi_0^{(e)}$ and
$\varphi_n^{(k)}=e^n\varphi_0^{(k)}$ are different from zero, then
they are eigenstates of, respectively, $X_-$ and $X_+$: \be\left\{
\begin{array}{ll}
X_-\,\varphi_n^{(e)}=-(n+1)\,\varphi_n^{(e)},\\
X_+\,\varphi_n^{(k)}=(n+1)\,\varphi_n^{(k)},
\end{array}\label{311}
\right.
 \en
for all $n=0,1,2,3,\ldots$.
\item It is clear from their definition that $X_\pm$ are not
expected to be positive or negative operators, even though
(\ref{311}) might suggest something different. Indeed this first
impression is correct, since it is also easy to continue the
analysis of the spectra of $X_\pm$ getting the following result: \be
\left\{
\begin{array}{ll}
X_+\,\varphi_n^{(k)}=(n+1)\,\varphi_n^{(k)}, \hspace{3mm}
n=0,1,2,\ldots\\
X_+\,\varphi_n^{(e)}=-n\,\varphi_n^{(e)}, \hspace{11mm}
n=0,1,2,\ldots\\
\end{array}\label{312}
\right.
 \en
as well as \be \left\{
\begin{array}{ll}
X_-\,\varphi_n^{(e)}=-(n+1)\,\varphi_n^{(e)}, \hspace{3mm}
n=0,1,2,\ldots\\
X_-\,\varphi_n^{(k)}=n\,\varphi_n^{(k)}, \hspace{17mm}
n=0,1,2,\ldots\\
\end{array}\label{313}
\right.
 \en
 It is clear that, since $(X_\pm)^\dagger\neq X_\pm$, there is no
 reason for all these different eigenstates to be mutually
 orthogonal, and in fact they are not. For the same reason, we
 can only conclude that  $\mathbb{Z}\subseteq \sigma(X_\pm)$, where
 $\sigma(X_\pm)$ are the spectra of the operators $X_+$ and $X_-$.
\item Using the explicit expressions for $e$ and $k$ we find that
\be X_+-\frac{1}{2}\; I=X_-+\frac{1}{2}\; I=
\frac{i}{2}\left\{(p_x-iy/2)^2+(p_y+ix/2)^2\right\},\label{314}\en
which shows that $-i\left(X_+-\frac{1}{2} I\right)$ and
$-i\left(X_-+\frac{1}{2} I\right)$ may be interpreted as a sort of
non-self adjoint Hamiltonian of a purely imaginary magnetic field
$\vec B_c$ arising from the following {\em complex vector potential}
$\vec A_c=\frac{i}{2}(-y, x, 0)$. This is amazing, because we
started with a Landau Hamiltonian with no magnetic field at all and
we have eventually recovered an imaginary and uniform $\vec B_c$.
The reason for this is related to the fact that the system in
question has a non-trivial geometry. In effect we are quantizing a
classical system living on the two dimensional plane with the origin
removed. The introduction of a vector potential with zero magnetic
field implies a gauge change which is reflected in the quantum
theory. The situation is reminiscent of the Bohm-Aharonov effect.
\end{itemize}
This is not yet the end of the story: other interesting operators
can still be defined starting from the ones we have considered
above. In particular, let us define \be
a=\frac{k+e^\dagger}{\sqrt{2}},\hspace{3mm}
a^\dagger=\frac{k^\dagger+e}{\sqrt{2}}, \,\Rightarrow [a,a^\dagger]=
I .\label{315} \en It is a simple exercise to check that
$a^\dagger\,a=H_0^\downarrow+\frac{1}{2}I$, where
$H_0^\downarrow={\frac 12}\,\left(p_x+{\frac y2}\right)^2+{\frac 12}\;
\left(p_y-{\frac x2}\right)^2$ differs from $H_0$ only through
the change of sign $\vec A\rightarrow -\vec A$, implying that  $\vec
B\rightarrow -\vec B$. Again, this result looks rather interesting:
although we started with a Hamiltonian for a free electron, the
introduction of a two-dimensional SUSY naturally {\em produced}
several operators, some self-adjoint, others not,  and describing real or
imaginary magnetic fields, yet whose spectra are analyzable in
great detail.

Of course the natural question, at this stage, is the following: is
it really SUSY that is  responsible for the appearance of $-\vec B$
in $H_0^\downarrow$?

\subsubsection{Case 2: $\partial_xW_2=\partial_yW_1=0$.}
%\label{III.2.2}

Let us consider again the commutation rules in (\ref{35}). What we
want to do now is to mimic, as far as possible, the standard Landau
level situation. This means, in particular, that we want $e, \,
e^\dag$ to commute with $k , \, k^\dag$. Therefore, because of
(\ref{35}), we need to have $\partial_xW_2=\partial_yW_1=0$ or, in
other words, the superpotential must have the following general
expression: $\vec W=(W_1(x),W_2(y),0)$. Needless to say, $\vec W_0$
satisfies this property, but it is also clear that this is not the
only possibility. Different choices produce, in general,
superpartner Hamiltonians which are really different, since
$\vec\nabla\cdot\vec W\neq 0$. The following results can be easily
deduced:
\begin{itemize}
\item if $\xi$ is an eigenstate of $h^\text{b}$ in (\ref{36}) with eigenvalue
$\epsilon$, then $e\xi$ is an eigenstate of $h^\text{f}$ with the
same eigenvalue. This is a standard result for partner Hamiltonians;

\item more interestingly, if we define the unitary operator
$T=e^{\overline{\alpha}k-\alpha k^\dagger}$, and we put
$\xi_n:=T^n\xi$, $n\in\mathbb{Z}$, it is also clear that $\xi_n$
is an eigenstate of $h^\text{b}$ with eigenvalue $\epsilon$ while
$a\xi_n$ is an eigenstate of $h^\text{f}$ again with the same eigenvalue.
This situation extends the analogous result valid for standard
Landau levels: once again, each {\em generalized Landau level} is
infinitely degenerate!
\end{itemize}

Thus, if we are able to generate superpotentials for which the
spectrum of $h^\text{b}$ is completely discrete we would be in the
standard SUSY situation and could build coherent states, using the
formalism presented above and generalizing (\ref{FQHE-VCS1}).

\subsubsection{Examples}

Our first choice of a superpotential $\vec W$ which is different
from the {\em standard one}, $\vec W_0$, is the following: $\vec
W=-\left(\dfrac{x+y}{2},\dfrac{x+y}{2},0\right)$. Note that with
this choice, even though $\partial_xW_2$ and $\partial_yW_1$ are
different from zero, in view of  (\ref{35}) we still have
$[e,e^\dagger]=[k,k^\dagger]= I$, $[k,e]=0$ and $[k,e^\dagger]= I$.
Therefore, $e$ and $k$ behave as a pair of {\bf coupled}
annihilation operators. However, using (\ref{39}), the magnetic
field associated to this $\vec W$ coincides with the one arising
from $\vec W_0$. So they describe the same physical situation.

A perhaps more interesting choice is the
superpotential, $\vec W= \kappa\left(\dfrac{1}{x},0,0\right)$, where
$\kappa$ is a constant to be conveniently determined later. Clearly, for this
potential $\partial_xW_2=\partial_yW_1=0$, so that  we are within the
framework  of {\bf Case 2} of  the previous subsection. With this
choice, let us introduce a slight change of notation, the reasons for which will become
clear shortly:
\be
  H^{\rm f} = h^{\rm b}, \quad H^{\rm b} = h^{\rm f}, \qquad A = e^{\dag}, \quad A^\dag = e\; .
\label{new-nation}
\en
Then,
\be
 H^{\rm f}= A A^\dag = \dfrac{1}{2}\left[p^2_x+\left(p_y -
 \dfrac{\kappa}{x}\right)^2-\dfrac{\kappa}{x^2}\right]=
 -\dfrac 12 \dfrac {\partial^2}{\partial x^2} + \dfrac {\kappa (\kappa  - 1 )}{2x^2}
  + \dfrac {i\kappa}x \dfrac \partial{\partial y} - \dfrac 12 \dfrac {\partial^2}{\partial y^2}\;.
 \label{newhamilt-f}
 \en
It is clear that $[H^{\rm f},p_y]=0$, so that the eigenstates of $H^{\rm f}$ can
be found among the eigenstates of the operator $p_y$. Consider the function
\be
 \Psi_{j m} (x, y ) = \psi (x) \chi_{j m}(x, y), \quad x, y  \in \mathbb R, \;\;
 j = 0, \pm 1, \pm 2, \ldots , \pm\infty, \;\;\; m = 0, 1, 2, 3, \ldots ,\infty,
 \label{ex2-hamiltonian}\en
where,
\be
 \chi_{jm}(x,y) = \left\{ \begin{array}{l} \displaystyle{\dfrac 1{\sqrt{2\pi}}}
  \exp \left( - i \displaystyle{\dfrac x{\vert x\vert}} my\right),
    \;\; \text{for} \;\; y \in [2j\pi ,\; 2(j+1)\pi ]\\[8pt]
    0, \;\; \text{otherwise} \end{array}   \right.
\label{ex2_soln}
\en
We then see that in order to obtain a solution to the eigenvalue problem
$H^{\rm f}\Psi_{j m} = \bvarepsilon^{\rm f}_{jm}\Psi_{j m}$, the function
$\psi(x)$ has to satisfy,
\be
\left[-\dfrac{1}{2}\,\dfrac{d^2}{dx^2} + \dfrac{\kappa m}{\vert x\vert}
+\dfrac{\kappa (\kappa -1)}{2x^2}
\right]\psi(x)=
\left(\bvarepsilon_{jm}^{\rm f} -\dfrac{m^2}{2}\right)\psi(x)\; .
\label{ex2-ev-eqn}
\en
Comparing this equation with the well-known radial equation for the
hydrogen atom:
\be
\left[ -\dfrac {\hbar^2}{2\mu}\;\dfrac{d^2 }{dr^2} - \dfrac {Ze^2}r +
\dfrac {\ell (\ell +1)\hbar^2}{2\mu r^2} \right]u = Eu\; ,
\label{hydrogen-eqn}
\en
we find  that for $m \neq 0$ and $\kappa = -1$ (\ref{ex2-ev-eqn}) reduces to
(\ref{hydrogen-eqn}), with the choice  $\ell =1,\; \dfrac {\hbar^2}\mu =1,\; Ze^2 = m$
and $E = \bvarepsilon^{\rm f}_{jm}-\dfrac{m^2}{2}$ and  if we
restrict $x$ to either $0\leq x < \infty$ or $-\infty < x \leq 0$. Explicitly, we then
get
\be
\left[-\dfrac{1}{2}\,\dfrac{d^2}{dx^2} - \dfrac{m}{\vert x\vert}
+\dfrac 1{x^2}
\right]\psi(x)=
\left(\bvarepsilon^{\rm f}_{jm}-\dfrac{m^2}{2}\right)\psi(x)\; .
\label{ex2-ev-eqn2}
\en
The solutions to (\ref{hydrogen-eqn}) come out in terms of the Laguerre polynomials
and the corresponding eigenvalues are
$$ E = - \dfrac {Z^2e^4 \mu}{2(n+ \ell +1)^2 \hbar^2},\qquad  n=0,1,2,3, \ldots \infty .$$
Hence the eigenvalues of $H^{\rm f}$ satisfy,
$$
  - \dfrac {m^2}{2( n+ 2)^2} = \bvarepsilon^{\rm f}_{jm}-\dfrac{m^2}{2}$$
whence, for $ j = 0, \pm 1 , \pm 2 , \pm\infty$,
\be
  \bvarepsilon^{\rm f}_{njm} := \bvarepsilon^{\rm f}_{jm} =  \dfrac{m^2}{2}\left[1 -
  \dfrac {1}{( n+ 2)^2}\right], \;\; m = 1,2,3, \ldots, \infty,
  \;\; n = 0,1,2, \ldots , \infty
  \;.
\label{ex2_evalues} \en Note that the eigenvalues $\bvarepsilon^{\rm
f}_{njm}$ do not depend on $j$ and hence each level, corresponding
to fixed values of $n$ and $m$, is infinitely degenerate. Moreover,
the lowest eigenvalue $\bvarepsilon^{\rm f}_{01m}$ is not zero:
$\bvarepsilon^{\rm f}_{01m} = \dfrac 38$.

  Let us next look at the other Hamiltonian, $H^{\rm b}$. From (\ref{38}) we easily
get,
\be
 H^{\rm b}= A^\dag A = \dfrac{1}{2}\left[p^2_x+\left(p_y - \dfrac{\kappa}{x}\right)^2-\dfrac{\kappa}{x^2}\right]=
 -\dfrac 12 \dfrac {\partial^2}{\partial x^2} + \dfrac {\kappa (\kappa  + 1 )}{2x^2}
  + \dfrac {i\kappa}x \dfrac \partial{\partial y} - \dfrac 12 \dfrac {\partial^2}{\partial y^2}\;.
 \label{newhamilt-b}
 \en
 Thus, taking $\kappa = -1$ and again assuming a solution of the type (\ref{ex2_soln}) and
taking note of (\ref{ex2-ev-eqn2}), we are
lead to the eigenvalue problem:
\be
\left[-\dfrac{1}{2}\,\dfrac{d^2}{dx^2} - \dfrac{m}{\vert x\vert} \right]\psi(x)=
\left(\bvarepsilon^{\rm b}_{jm}-\dfrac{m^2}{2}\right)\psi(x)\; .
\label{ex2-ev-eqn4}
\en
Comparing with the equation for the hydrogen atom, (\ref{hydrogen-eqn}), we see that
we are in the case where $\ell =0$. Thus, analogously to (\ref{ex2_evalues}) we get,
for $ j = 0, \pm 1 , \pm 2 , \pm\infty$,
\be
  \bvarepsilon^{\rm b}_{njm} := \bvarepsilon^{\rm b}_{jm} =  \dfrac{m^2}{2}\left[1 -
  \dfrac {1}{( n+ 1)^2}\right], \;\; m = 1,2,3, \ldots, \infty,
  \;\; n = 0,1,2, \ldots , \infty
  \;.
\label{ex2_evalues4}
\en
Note that, as expected,
\be
  \bvarepsilon^{\rm f}_{njm} = \bvarepsilon^{\rm b}_{(n+1)jm}\; .
\label{ener-reln}
\en
This time, the lowest eigenvalue, coming at $n=0$, is in fact zero, which justifies the
change in the identification of the bosonic and fermionic sectors in (\ref{new-nation}).
 Moreover, this eigenvalue is doubly degenerate, i.e.,
\be
   \bvarepsilon^{\rm b}_{0jm} = 0 , \quad m=1,2,3, \ldots , \infty, \quad
      j = 0, \pm 1, \pm 2, \ldots , \pm\infty\; .
\label{ground-ev-degen}
\en
Finally, a straightforward computation (or an inspection of the well-known ground state radial wave
functions for the hydrogen atom) yields for the eigenstate, $\Psi_{0jm}(x,y)$ corresponding to the
eigenvalue $\bvarepsilon^{\rm b}_{0jm} = 0$, the function,
\be
\Psi_{0jm}(x,y) = N\vert x\vert e^{-m\vert x\vert}\;\chi_{jm} (x,y),
\label{ground-state}
\en
with $\chi_{jm}$ as in (\ref{ex2_soln}) and $N$ being a normalization constant.
Finally, it is easily checked that $A\Psi_{0jm} = 0$.

  It is now possible to construct VCS for the supersymmetric pair $\{H^{\rm b}, \; H^{\rm f}\}$,
for fixed values of $m$ and $j$. Let $\Psi_{njm}^{\rm b}$ be the eigenvectors of the Hamiltonian
$H^{\rm b}$, corresponding to the eigenvalues $\bvarepsilon^{\rm b}_{njm}$ and let $\h_{jm}^{\rm b}$
be the Hilbert space generated by the vectors $\Psi_{njm}^{\rm b}, \; n=0,1,2, \ldots, \infty$.
Similarly, let $\Psi_{njm}^{\rm f}$ be the eigenvectors of the Hamiltonian
$H^{\rm f}$, corresponding to the eigenvalues $\bvarepsilon^{\rm f}_{njm}$ and  $\h_{jm}^{\rm f}$
the Hilbert space generated by the vectors $\Psi_{njm}^{\rm f}, \; n=0,1,2, \ldots, \infty$. Set
$\h^{\rm SUSY}_{jm} = \h_{jm}^{\rm b} \oplus \h_{jm}^{\rm f}$. Then, following (\ref{VCS-2comp}), we
define vector coherent states on $\h^{\rm SUSY}_{jm}$ as
\be
  \vert z, \overline{z};\; jm\rangle =
  \mathcal N(\vert z\vert^2)^{-\frac 12}\sum_{n=0}^\infty \; \begin{pmatrix}
   \displaystyle \frac {z^n}{\sqrt{\bvarepsilon_{njm}^{\rm b} !}}\Psi_{njm}^{\text{b}}\\[20pt]
    \displaystyle\frac {\overline{z}^{n+1}}{\sqrt{\bvarepsilon_{(n+1)jm}^{\rm b} !}}
    \Psi_{njm}^{\text{f}}
    \end{pmatrix}, \quad \mathcal N(\vert z\vert^2) = 1 +2\sum_{n=1}^\infty
       \frac {\vert z\vert^2}{\bvarepsilon_{njm}^{\rm b} !}\; .
\label{VCS-ex2}
\en
In order to get a resolution of the identity, we note that the radius of convergence of
the series representing $N(\vert z\vert^2)$ is $\dfrac {m^2}2$ and furthermore,
\be
\bvarepsilon_{njm}^{\rm b} ! = \frac {m^{2n}}{2^{n+1}}\;\left[ 1 + \frac 1{n+1}\right].
\label{factorial}
\en
Thus following (\ref{mom-prob1}), we need to find a measure $d\lambda$ such that
\be
  2\pi\int_0^{\frac m{\sqrt{2}}}r^{2n}\; d\lambda =
                  \frac {m^{2n}}{2^{n+1}}\;\left[ 1 + \frac 1{n+1}\right]\; .
\label{mom-prob-ex2}
\en
The measure in question is easily found to be (see also \cite{gazklau} for a similar
computation)
\be
 d\lambda(r) = \frac 1{4\pi} \delta (r - \tfrac m{\sqrt{2}}^- )\; dr +
 \frac 1{\pi m^2} r\; dr\; .
\label{resoli-id-meas-ex2}
\en
Thus, there follows the resolution of the identity,
\be
\int_0^{\frac m{\sqrt{2}}}\!\!\int_0^{2\pi}\vert z, \overline{z};\; jm\rangle
\langle z, \overline{z};\; jm\vert\;\mathcal N(\vert z\vert^2)\;d\lambda (r)\; d\theta =
\begin{pmatrix} I_{\h_{jm}^{\rm b}} & 0 \\ 0 & I_{\h_{jm}^{\rm f}} \end{pmatrix}
= I_{\h_{jm}^{\rm SUSY}}.
\label{resolid-ex2}
\en
(where $z = re^{i\theta}$). Also, in view of the fact that
$$\bvarepsilon_{njm}^{\rm b} ! = \frac {2^{n+1}}{m^{2n}}\left[ 1 - \frac 1{n+2}\right], $$
the normalization factor $\mathcal N(\vert z\vert^2)$ is computed to
be, \be
 \mathcal N(\vert z\vert^2) = \frac {4u}{1-u} - \frac {u^2}4\log (1-u) -3,
 \qquad u = \frac {2\vert z\vert^2}{m^2}\; .
 \label{norm-const-ex2}
 \en

\vspace{2mm}

Of course there could be other choices of $\vec W$ as well, producing
other families of   VCS.  For example, $\vec
W=\left(-\dfrac{x^2}{2},0,0\right)$ could be one such interesting choice.
For this superpotential the
operator $h^{\rm b}$ looks like $$h^{\rm b} =\dfrac{1}{2} \left(p^2_x+
p_y^2+x^2p_y+ \dfrac{x^4}{4}-x\right),$$ which again commutes with
$p_y$, and with the choice of a trial solution similar to (\ref{ex2-hamiltonian}),
it is easy to deduce the following one-dimensional eigenvalue equation:
\be
\dfrac{1}{2}\left(-\dfrac{d^2}{dx^2}+\dfrac{x^4}{4}+x^2k-x\right)\psi(x)
=\left(E-\dfrac{k^2}{2}\right)\psi(x),
\label{es1}
\en
where as before
$k\in\mathbb{Z}$. Solving this equation is harder than the previous
one. However, it is an easy exercise to find the ground state $\psi_0$
which, as before, turns out to be infinitely degenerate.  The equation for $\psi_0$ is
$e\psi_0 = 0$, whose solution is
$\psi_0 (x,y)=\dfrac{N}{\sqrt{2\pi}}\,e^{-kx+iky-x^3/6}$. Again, it is
easy to check that for each $k\in\mathbb{Z}$, this satisfies the
equation $h^{\rm b}\psi_0 =0$.

Finding the excited eigenstates is a more difficult  problem this time, and
one can look for solutions of (\ref{es1}) in power series. In this
way one can get a (formal) solution, which, however, does not appear
to be in  closed form. Next, these (formal) eigenstates, can be used to find the
eigenstates of the hamiltonian $h^{\rm f}$, using
(\ref{ladder_op1}). Subsequently,  using (\ref{VCS1}), one can again
construct VCS.

\section{Conclusions}

As mentioned in the Introduction,  we have presented in this paper a
method for constructing vector coherent states for supersymmetric
Hamiltonian pairs and then applied it to constructing such states to
pairs of Hamiltonians arising from a generalization of the
fractional quantum Hall effect. While the general scheme adopted
here for constructing VCS has been developed elsewhere, the
application to supersymmetric Hamiltonians is new. Two interesting
facts ought to be reiterated here. The first is the appearance of
both analytic and anti-analytic functions in the complex
representation of the underlying Hilbert space, in which the bosonic
part occupies the analytic and the fermionic part the anti-analytic
sectors. The second is the fact that this complex representation is
naturally equivalent to a representation using the standard
anti-commuting Grassmann variables, common to treatments of
supersymmetry.

As a concrete example of our construction we have considered the
Hamiltonian of the Landau levels and some natural generalizations
of it, suggested by SUSY: this in turn, produced several SUSY partner
Hamiltonians and, as a consequence, their related VCS.

\section*{Acknowledgements}

This work was partially supported by the Ministero Affari Esteri,
Italy, through its program of financial support for international
cooperations, Bando CORI 2003, cap. B.U. 9.3.0001.0001.0001, and
through grants from the Natural Sciences and Engineering Research
Council (NSERC), Canada and the Fonds qu\'eb\'ecois de la recherche
sur la nature et les technologies (FQRNT), Qu\'ebec. One of us (STA)
would like to thank H. Upmeier for useful suggestions.


\begin{thebibliography}{99}

\bibitem{alienggaz} S.T. Ali, M. Engli\v s and J.-P. Gazeau, {\em Vector coherent
states from Plancherel's Theorem and Clifford algebras\/}, J. Phys. {\bf A37}, 6067-6089 (2004).

\bibitem{ab} S.T. Ali and F. Bagarello, {\em Some physical appearances of
vector coherent states and coherent states related to
 degenerate Hamiltonians}, J. Math. Phys. {\bf 46}, 053518 (2005).

\bibitem{b6} J.P. Antoine, F. Bagarello,  {\em
Localization properties and wavelet-like orthonormal bases for the
lowest Landau level}, in {\em Advances in Gabor Analysis}, H.G.
Feichtinger, T. Strohmer, Eds., Birkh\"auser, Boston, 2003.

\bibitem{b1} J.P. Antoine, F. Bagarello, {\em Wavelet-like
orthonormal basis of the lowest Landau level}, J. Phys. A {\bf
27}, 2471-2481 (1994).

\bibitem{bag2005} F. Bagarello, {\em Relations between multi-resolution
analysis and quantum mechanics},  J. Math. Phys. {\bf 46},
 053506 (2005).

\bibitem{b5} F. Bagarello {\em Multi-resolution analysis and fractional
quantum Hall effect: more results}, J. Phys. A {\bf 36}, 123-138 (2003).

\bibitem{b7} F. Bagarello {\em Multi-resolution analysis
generated by a seed function},  J. Math. Phys., {\bf 44},
1519-1534 (2003).

\bibitem{b4} F. Bagarello {\em
Multi-resolution analysis and fractional quantum Hall effect: an
equivalence result}, J. Math. Phys., {\bf 42}, 5116-5129, (2001).

\bibitem{b3} F. Bagarello, {\em Applications of
wavelets to quantum mechanics: a pedagogical example}, J. Phys. A
{\bf 29}, 565-576 (1996).

\bibitem{b2} F. Bagarello, {\em More
wavelet-like orthonormal bases for the lowest Landau level:  Some
considerations}, J. Phys. A {\bf 27}, 5583-5597 (1994).

\bibitem{bms} F. Bagarello, G. Morchio and F. Strocchi, {\em Quantum
corrections to the
Wigner crystal.  An Hartree-Fock expansion}, Phys. Rev. B {\bf
48}, (1993), pp. 5306.

\bibitem{borklilesri} D. Borthwick, S. Klimek, A. Lesniewski and M. Rinaldi,
{\em Super Toeplitz operators and non-perturbative deformation quantization of
supermanifolds\/,} Commun. Math. Phys., {\bf 153}, 49-76 (1993).

\bibitem{daspern} A. Das and S.A. Pernice, {\em Higher dimensional SUSY
quantum mechanics}, Mod. Phys. Lett., A {\bf 12},  581-588 (1997).

\bibitem{fatkosnietru} B.W. Fatyaga, V.A. Kosteleck\'y, M.M. Nieto and D.R. Truax,
{\em Supercoherent states\/,} Phy. Rev. D., {\bf 43}, 1403-1412 (1991).

\bibitem{ferhunie} D.J. Fern\'andez, V. Hussin and L.M. Nieto, {\em Coherent states
for isospectral oscillator Hamiltonians\/,} J. Phys. A., {\bf 27}, 3547-3564 (1994).

\bibitem{gazklau} J.-P. Gazeau and J.R. Klauder, {\em Coherent States for Systems
with Discrete and Continuous Spectrum\/}, J. Phys. A., {\bf 32}, 123-132 (1999).

\bibitem{junker} G. Junker, {\em Supersymmetric Methods in Quantum and Statistical
Physics\/,} Springer, Berlin (1996)

\bibitem{juroy} G. Junker and P. Roy, {\em Non-linear coherent states associated to
conditionally exactly solvable problems\/,} Phys. Lett. A., {\bf 257}, 113-119 (1999).

\bibitem{loaup} M. Loaiza and H. Upmeier, {\em Toeplitz $C^*$-algebras on super Cartan domains\/,}
preprint, University of Marburg (2007).

\bibitem{metz} H. Maier-Metz, {\em Supersymmetry in Landau levels},
 Eur. J. Phys., {\bf 19}, 137-141 (1998)

\bibitem{mal} S.G. Mallat, {\em Multiresolution approximations and
wavelets orthonormal bases of
$\Lc^2(\R)$}, Trans. Am. Math. Soc. {\bf 315}, Nr. 1, 69-87
(1989).

\bibitem{thirali} K. Thirulogasanthar and S.T. Ali, {\it A class  of vector coherent states
defined over matrix domains\/,} J. Math. Phys. {\bf 44}, 5070-5083 (2003).



\end{thebibliography}
\end{document}